\documentclass{article}
\usepackage{slashed,feynmf,epsfig,amsmath,amssymb,enumitem}
\usepackage[papersize={8.5in,11in}]{geometry}
\renewcommand{\vec}[1]{\boldsymbol{\mathrm{#1}}}
\usepackage{graphicx}
\newcommand{\be}{\begin{equation}}
\newcommand{\ee}{\end{equation}}
\begin{document}
\title{Gravitational Theory of Cosmology, Galaxies and Galaxy Clusters}
\author{J. W. Moffat\\~\\
Perimeter Institute for Theoretical Physics, Waterloo, Ontario N2L 2Y5, Canada\\
and\\
Department of Physics and Astronomy, University of Waterloo, Waterloo,\\
Ontario N2L 3G1, Canada}
\maketitle




\begin{abstract}
A modified gravitational theory explains early universe and late time cosmology, galaxy and galaxy cluster dynamics. The modified gravity (MOG) theory extends general relativity (GR) by three extra degrees of freedom: a scalar field $G$, enhancing the strength of the Newtonian gravitational constant $G_N$, a gravitational, spin 1 vector graviton field $\phi_\mu$, and the effective mass $\mu$ of the ultralight spin 1 graviton. For $t < t_{\rm rec}$, where $t_{\rm rec}$ denotes the time of recombination and re-ionization, the density of the vector graviton $\rho_\phi > \rho_b$, where $\rho_b$ is the density of baryons, while for $t > t_{\rm rec}$ we have $\rho_b > \rho_\phi$. The matter density is parameterized by $\Omega_M=\Omega_b+\Omega_\phi+\Omega_r$ where $\Omega_r=\Omega_\gamma+\Omega_\nu$. For the cosmological parameter values obtained by the Planck Collaboration, the CMB acoustical oscillation power spectrum, polarization and lensing data can be fitted as in the $\Lambda$CDM model. When the baryon density $\rho_b$ dominates the late time universe, MOG explains galaxy rotation curves, the dynamics of galaxy clusters, galaxy lensing and the galaxy clusters matter power spectrum without dominant dark matter.
\end{abstract}

\maketitle

\section{Introduction}

Dark matter was introduced to explain the stable dynamics of galaxies and galaxy clusters. General relativity (GR) with only ordinary baryon matter cannot explain the present accumulation of astrophysical and cosmological data without dark matter. However, dark matter has not been observed in laboratory experiments~\cite{Baudis}. Therefore, it is important to consider a modified gravitational theory. The observed acceleration of the universe has also complicated the situation by needing a dark energy, either in the form of the cosmological constant vacuum energy or as a modification of GR.

The fully relativistic and covariant modified gravity (MOG) theory Scalar-Tensor-Vector-Gravity (STVG)\\
~\cite{Moffat,BrownsteinMoffat1,BrownsteinMoffat2,BrownsteinMoffat3,Brownstein,MoffatToth1,MoffatToth2,MoffatToth3,MoffatToth4,MoffatRahvar1,MoffatRahvar2,
GreenMoffat,Roshan1,Roshan2} has been successfully applied to explain the rotation curves of galaxies and the dynamics of galaxy clusters. In addition to the enhanced gravitational scalar field strength $G$, the theory also has a repulsive gravitational field generated by a massive, spin 1 Proca vector graviton (VG) field $\phi_{\mu}$. The mass of the VG is determined by a scalar field $\mu$. The value of $\mu$ that fits the galaxy rotation curves and the cluster dynamics is $\mu\sim 0.01 - 0.04\,{\rm kpc}^{-1}$, corresponding to $\mu^{-1}\sim 25 - 100$ kpc and a mass $m_\phi\sim ~ 10^{-28}\,{\rm eV}$. The mass of the VG in the early universe is $m_\phi\sim 10^{-22}\,{eV}$. The massless spin 2 graviton associated with $g_{\mu\nu}$ and the massive VG are difficult and probably impossible to detect~\cite{Dyson,Rothman}. This will explain the so far lack of success of detecting a dark matter particle.

The difference between standard dark matter models and MOG is that dark matter models assume that GR is the correct theory of gravity and a dark matter particle such as WIMPS, axions and fuzzy dark matter~\cite{Witten} are postulated to belong to the standard particle model. MOG extends GR by the addition of new gravitational degrees of freedom. The cosmological observations attributed to undetected dark matter particles are explained as being of {\it purely gravitational origin}. The gravitational coupling of the VG to matter is universal with the gravitational charge $Q_g=\sqrt{8\pi\alpha G_N}M=\kappa M$, where $\alpha$ is a dimensionless parameter associated with the scalar coupling $G=G_N(1+\alpha)$, $M$ is the mass of a body, $G_N$ is Newton's gravitational constant. Moreover, $\kappa = {\cal O}(1/M_{PL})$ where $M_{\rm PL}= 1/\sqrt{8\pi G_N}=2.435\times 10^{18}$ GeV is the reduced Planck mass. As the universe expands to the present epoch, the VGs become ultra-relativistic and cannot form bound states such as galaxy and galaxy cluster halos. MOG becomes the dominant gravitational phase with the density of baryons $\rho_b$ exceeding the density $\rho_\phi$ of the VGs with the decreased mass $m_\phi\sim 10^{-28}$ eV. Our formulation of gravitation fits the cosmological data described by the $\Lambda$CDM model, as well as the present epoch galaxy and galaxy cluster data without dominant dark matter.

A complete description of the universe by a theory requires not only that the present day observations of galaxies and galaxy clusters require an explanation, but also that the wealth of cosmological observations is described as well by the theory. For MOG there are two possible approaches to an explanation of the cosmological data. One is to pursue a description without dark matter employing only baryonic matter and the baryonic density $\rho_b$. This approach has been explored with some success~\cite{MoffatToth3,MoffatToth4,Amendola} but requires further exploration in the future. The second approach is to promote the gravitational VG (spin 1 vector graviton) degree of freedom as a dominant density $\rho_\phi$ in the early universe, so that it dominates the baryon density $\rho_b$ for the time before the reionization phase when stars and galaxies are first formed~\cite{Moffat,Moffat2}. After this time, there is a transition to modified gravity when $\rho_b > \rho_\phi$. This transition can be caused by a matter-energy phase transition. In the following, we present a further investigation of the second approach.

\section{The Friedmann Equations}

We base our cosmology on the homogeneous and isotropic Friedmann-Lemaitre-Robertson-Walker (FLRW) background metric:
\begin{equation}
ds^2=dt^2-a^2(t)\biggl[\frac{dr^2}{1-Kr^2}+r^2(d\theta^2+\sin^2\theta d\phi^2)\biggr],
\end{equation}
where $K=-1,0,1\,({\rm length}^{-2})$ for open, flat and closed universes, respectively. We use the energy-momentum tensor of a perfect fluid in (\ref{energymom}).

We have
\begin{equation}
\label{rho density}
\rho=\rho_b+\rho_\phi+\rho_G+\rho_\mu+\rho_r,
\end{equation}
where $\rho_b,\rho_\phi, \rho_G$ and $\rho_\mu$ denote the densities of baryon matter, the electrically neutral VG field $\phi_\mu$, the scalar $G$ field, the effective mass $\mu$ of the VG and radiation, respectively. The radiation density $\rho_r=\rho_\gamma+\rho_\nu$, where $\rho_\gamma$ and $\rho_\nu$ denote the densities of photons and neutrinos, respectively.

Due to the symmetries of the FLRW background spacetime, we have $\phi_0\neq 0,\phi_i\sim 0\,(i=1,2,3)$ and $B_{\mu\nu}\sim 0$. We have from (\ref{Bequation1}) and (\ref{current}):
\be
\rho\sim\frac{\mu^2}{\kappa}\phi_0,
\ee
where $\rho$ is given by (\ref{rho density}) and
\be
{T_{0\phi}}^0=\rho_\phi=\frac{1}{8\pi}\mu^2\phi_0^2.
\ee
The kinetic energy of the VGs is small, $B^{\mu\nu}\sim 0$, because the VGs are close to the ground state energy of a Bose-Einstein condensate superfluid.

In the following, we assume a spatially flat universe $K=0$ and $V_G=V_\mu=0$. Furthermore, we assume that the time dependence
of the gravitational field strength is negligible, $\dot G\sim 0$, and we also assume that $\dot\mu\sim 0$. We obtain the approximate Friedmann equations:
\begin{equation}
\label{Friedmann3}
\biggl(\frac{\dot a}{a}\biggr)^2=\frac{8\pi G\rho}{3}+\frac{\Lambda}{3},
\end{equation}
\begin{equation}
\frac{\ddot a}{a}=-\frac{4\pi G}{3}(\rho+3p)+\frac{\Lambda}{3}.
\end{equation}
The energy conservation equation is
\begin{equation}
\dot\rho+3\frac{d\ln a}{dt}(\rho+p)=0.
\end{equation}
For the matter dominated universe with $p=0$ we have $\rho_b\propto 1/a^3$.

We introduce comoving coordinates:
\begin{equation}
{\bf x}=\frac{a_0{\bf r}}{a(t)}.
\end{equation}
The Fourier expanded density, pressure and gravitational potential in terms of plane waves with comoving wavenumber ${\bf k}$ are given by
\begin{equation}
\delta\rho({\bf r,t})=\frac{1}{(2\pi)^{3/2}}\bar\rho\int d^3k\delta_{\bf k}(t)\exp(i(a_0/a){\bf k}\cdot{\bf r}),
\end{equation}
\begin{equation}
\delta p({\bf r,t})=\frac{1}{(2\pi)^{3/2}}\int d^3k\delta p_{\bf k}(t)\exp(i(a_0/a){\bf k}\cdot{\bf r}),
\end{equation}
\begin{equation}
\Phi({\bf r,t})=\frac{1}{(2\pi)^{3/2}}\int d^3k\Phi_{\bf k}(t)\exp(i(a_0/a){\bf k}\cdot{\bf r}),
\end{equation}
where $\delta=\delta\rho/{\bar\rho}$ is the relative density perturbation contrast. We have for the gravitational potential:
\begin{equation}
k^2\Phi_{\bf k}=-4\pi G\biggl(\frac{a}{a_0}\biggr)^2\bar\rho\delta_{\bf k},
\end{equation}
where the MOG potential for a given density $\rho({\vec r})$ is
\be
\label{MOGpotential}
\Phi(\vec{r}) = - G_N \int d^3{\vec r}'\frac{\rho(\vec{r}')}{|\vec{r}-\vec{r}'|}\Big[1+\alpha -\alpha e^{-\mu|\vec{r}-\vec{r}'|}\Big].
\ee
The MOG acceleration equation is given by
\begin{equation}
\label{MOG acceleration}
a_{\rm MOG}({\vec r})=-G_N\int d^3{\vec r}'\frac{\rho({\vec r}')({\vec r}-{\vec r}')}{|{\vec r}-{\vec r}'|^3}
[1+\alpha-\alpha\exp(-\mu|{\vec r}-{\vec r}'|(1+\mu|{\vec r}-{\vec r}'|))].
\end{equation}

The density of VGs in the early universe is given by
\be
\label{phi density}
\Omega_\phi\sim 0.12\biggl(\frac{\kappa}{10^{18}\,{\rm GeV}}\biggr)^2\biggl(\frac{m_\phi}{10^{-22}\,{\rm eV}}\biggr)^{1/2}.
\ee
The energy density $\rho_\phi$ scales as $1/a^3$ just like matter in the matter dominated universe. The mean velocity of the VGs is given by
\be
\label{VG velocity}
{\bar v}\sim\frac{2\pi\hbar}{\lambda m_\phi}\sim 10\,{\rm kms^{-1}}\biggl(\frac{10^{-22}\,{\rm eV}}{m_\phi}\biggr),
\ee
where the VG de Broglie wave length $\lambda/2\pi\sim \hbar/m_\phi{\bar v}$.

\section{MOG Structure Growth and Angular Acoustical Power Spectrum}

We assume that when $\rho_\phi$ dominates in the early universe, $\alpha < 1$ and $G\sim G_N$.
The particle densities are expressed as the ratios $\Omega_x=8\pi G_N\rho_x/3H^2$. In particular, we have for the baryon, VGs and the cosmological constant $\Lambda$:
\begin{equation}
\label{Omegaeqs}
\Omega_b=\frac{8\pi G_N\rho_b}{3H^2},\quad\Omega_\phi=\frac{8\pi G_N\rho_\phi}{3H^2},\quad\Omega_{\Lambda}=\frac{\Lambda}{3H^2}.
\end{equation}
At the time of big-bang nucleosynthesis (BBN), we have $G\sim G_N$, guaranteeing that the production of elements at the time of BBN agrees with observations.
We assume that at horizon entry until some time after decoupling, $\rho_\phi\gg \rho_b$, $\rho_\phi\gg\rho_G$, $\rho_\phi\gg\rho_\mu$, so that $\rho\sim \rho_b+\rho_\phi+\rho_r$~\cite{Moffat,Moffat2}.

The equation for the density perturbation for the non-relativistic single-component fluid is given by the Jeans equation:
\begin{equation}
\label{deltaeq}
\ddot\delta_{\bf k}+2H\dot\delta_{\bf k}+\bigg(\frac{c_s^2a_0^2k^2}{a^2}-4\pi G_N\bar\rho\biggr)\delta_{\bf k}=0,
\end{equation}
where $H={\dot a}/a$ and $\bar\rho$ denotes the mean density. Here, $c_s$ is the speed of sound:
\begin{equation}
c_s=\sqrt{\frac{dp}{d\rho}},
\end{equation}
and perturbations with $\delta p/\delta\rho=(d\bar {p}/dt)/(d\bar{\rho}/dt)=c_s^2$ are called adiabatic perturbations and for these perturbations $p=p(\rho)$. The first term in the parenthesis of (\ref{deltaeq}) is due to the adiabatic perturbation pressure contribution $\delta p=c_s^2\delta\rho$.

The nature of the solutions to (\ref{deltaeq}) depends on the sign of the factor in parenthesis. Pressure attempts to resist compression, so when the pressure term dominates, we obtain an oscillatory solution comprising standing density (sound) waves. The second term is due to gravity and when this term dominates the perturbations grow. The Jeans wavenumber when the pressure and gravity terms are equal is given by
\begin{equation}
k_J=\frac{a}{a_0}\frac{\sqrt{4\pi G_N\bar\rho}}{c_s},
\end{equation}
corresponding to the wavelength $\lambda_J=2\pi/k_J$ (Jeans length) and for non-relativistic matter $k_J\gg {\cal H}=(a/a_0)H$ and $c_s\ll 1$ where ${\cal H}$ is the comoving Hubble scale.

The Proca vector field $\phi_\mu$ is a neutral massive VG. Because the VGs are electrically neutral, they do not couple to photons and {\it they can be treated as almost pressureless}, so the pressure gradient term in the Jeans equation is absent and $c_s\sim 0$. The Jeans length is approximately zero and we have
\begin{equation}
\label{phipertequation}
\ddot\delta_{\bf k}+2H\dot\delta_{\bf k}-4\pi G_N\bar\rho\delta_{\bf k}=0.
\end{equation}
For the radiation dominant era $\rho_\phi < \rho_r$, but perturbations in radiation, $\delta\rho_r$, oscillate rapidly and are damped $\delta\rho_r\sim 0$, so they can be ignored compared to the $\delta\rho_\phi$ perturbations. The expansion law is that for a radiation dominated universe:
\begin{equation}
a(t)\propto t^{1/2},\quad H=\frac{1}{2t}=\sqrt{\frac{8\pi G_N}{3}\bar\rho}.
\end{equation}
Dividing (\ref{phipertequation}) by $H^2$ we get
\begin{equation}
\frac{\ddot\delta_{\bf k}}{H^2}+2\frac{\dot\delta_{\bf k}}{H}=\frac{3}{2}\frac{\bar\rho_\phi}{\bar\rho}\delta_{\bf k}.
\end{equation}
The right-hand-side can be dropped because $\bar\rho_\phi < \bar\rho$. The solution to this equation is given by
\begin{equation}
\delta_{\bf k}=a+b\ln t,
\end{equation}
where $a$ and $b$ are constants. We find that the perturbations due to the VGs grow at most logarithmically during the radiation dominated era of the universe. The increased expansion rate due to the presence of the smooth radiation component slows down the growth of perturbations.

Because the VGs are almost pressureless and the speed of sound $c_s\sim 0$, the VG Jeans length is approximately zero, so all scales are larger than the Jeans scale, and there is no oscillatory behavior for the VG particles. Instead, perturbations grow at all length scales. On the other hand, the $\rho_{b\gamma}$ perturbations oscillate before decoupling. The dark energy determined by the cosmological constant $\Lambda$ is a smooth component of energy and only becomes important at later times.

Due to decoupling the baryon Jeans length is comparable to the Hubble length, so adiabatic baryon perturbations will oscillate before decoupling. The baryon density perturbation begins to grow only after the decoupling time $t=t_{dec}$, because before decoupling the baryon-photon pressure prevents any growth. Without the VG density $\rho_\phi$ the baryon perturbations grow as $\delta_b\propto a\propto t^{2/3}$ after decoupling, and the baryon density perturbations $\delta\rho_b$ catch up to the VG density perturbations $\delta\rho_\phi$. However, the CMB anisotropy due to baryon density perturbations observed at $t=t_{dec}$ are too small $\sim 10^{-4}$ to produce a growth factor of 1100 needed to generate the presently observed large-scale structure. The VG density perturbations solve the problem, for they begin to grow earlier at horizon entry and by $t=t_{dec}$ they are significantly greater than the baryon density perturbations.

The baryon sound wave oscillations due to the baryon-photon pressure prior to the decoupling time produce acoustical peaks in the angular power spectrum, ${\cal D}_\ell=\ell(\ell+1)C_\ell/2\pi$. Because the form of the MOG Friedmann equation (\ref{Friedmann3}) is the same as in the ${\Lambda}CDM$ model, we can match the $\Lambda$CDM calculation of the CMB angular power spectrum. The calculation of the CMB power spectrum is displayed in Fig.1. The calculation of the power spectrum in the ${\Lambda}CDM$ model is duplicated in MOG, using the Planck 2018 Collaboration best-fit parameter values~\cite{Planck}: $\Omega_bh^2=0.0224\pm 0.0001$, $\Omega_\phi h^2=0.120\pm 0.001$, $\Omega_\Lambda=0.680\pm 0.013$, $n_s=0.965\pm 0.004$, $\sigma_8=0.811\pm 0.006$, $H_0=67.4\pm 0.5\,{\rm km}\,{\rm sec}^{-1}\,{\rm Mpc}^{-1}$, together with the remaining parameters in the fitting process.
\
\begin{figure}
\centering \includegraphics[scale=1.0]{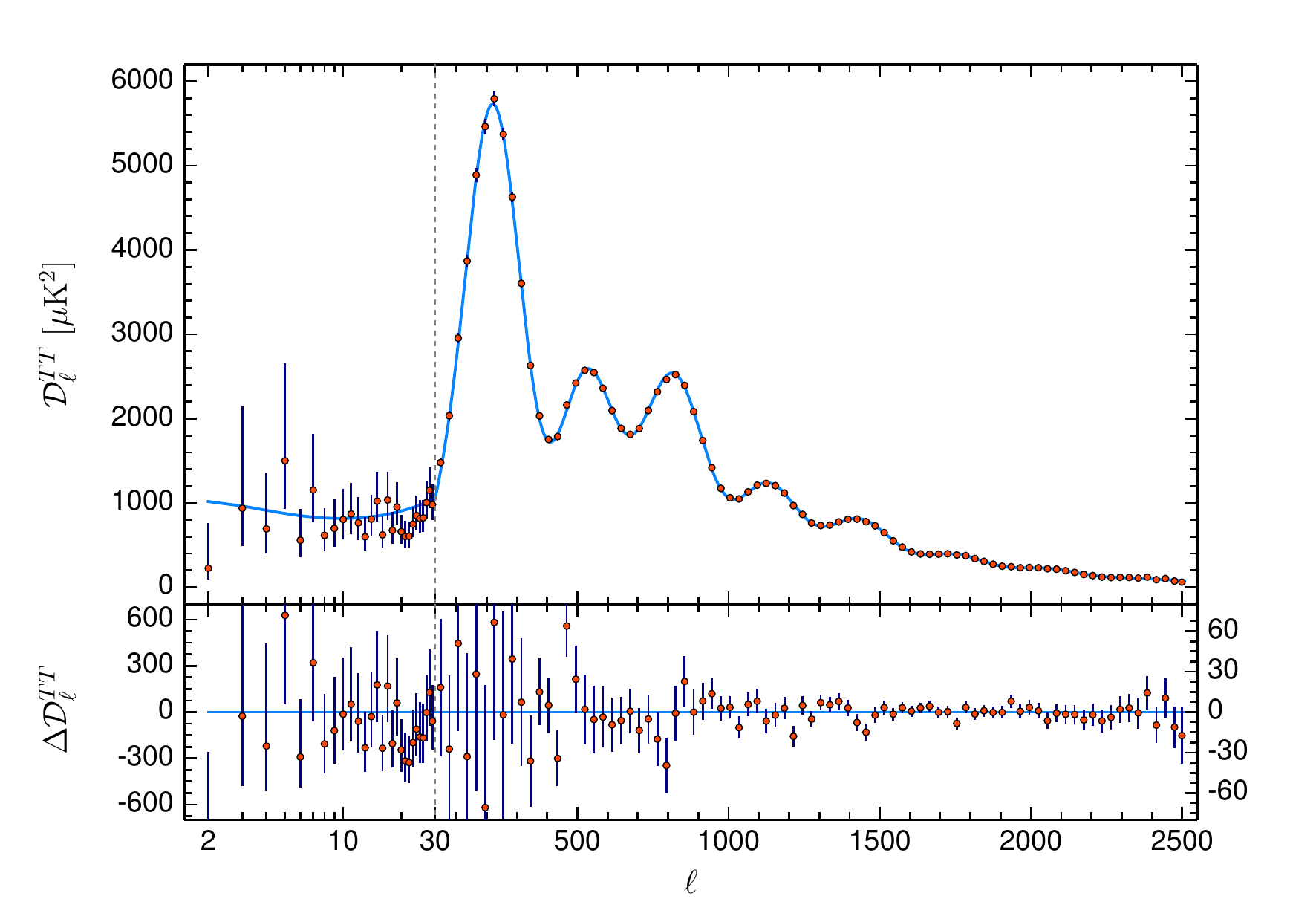}\\
\caption{\label{fig:angularpowerspectrum}The TT power spectrum obtained from ref.~\cite{Planck}, using the best-fit parameters of the Planck Collaboration~\cite{Planck}. The red line shows the best-fit MOG spectrum. The lower panel shows the residuals with respect to the $\Lambda$CDM model and the MOG model.}
\end{figure}

\section{Late Time Universe}

As the universe expands beyond the time of decoupling, the gravitational attraction between the baryons becomes enhanced as the parameter $\alpha$ increases in size and $G=G_\infty=G_N(1+\alpha)$. The density $\rho_\phi$ and the increasing enhancement of the size of $G$ deepen the baryon gravitational potential well. Eventually, as the large scale structures form $\rho_\phi < \rho_b$ and the baryon dominated MOG dynamics takes over. Once the combined $\rho_b$ and $\rho_\phi$ density perturbations have grown sufficiently to produce stars and galaxies at about 400 million years after the big bang, the MOG non-relativistic dynamics for baryons takes over to determine the final evolution and dynamics of galaxies and galaxy clusters in the present epoch. The effective mass $\mu$ undergoes a decrease, and from the best-fit values from galaxy rotation curves and cluster dynamics~\cite{MoffatRahvar1,MoffatRahvar2,GreenMoffat}, $\mu=0.01 - 0.04\,{\rm kpc}^{-1}$, corresponding to $m_\phi\sim 10^{-28}\,{\rm eV}$. 

From (\ref{phi density}) for a VG mass $m_\phi\sim 10^{-28}$ eV, the VG density in the present universe is $\Omega_\phi \leq 10^{-3}$. From (\ref{VG velocity}) the mean velocity of the VGs is now of order the speed of light $c$ and the VGs are ultra-relativistic particles, so galaxy and galaxy cluster halos of VGs cannot form as stable bound systems. The de Broglie wavelength, in natural units, $\lambda_\phi\sim 1/m_\phi$, of the ultra-light VGs is of the size of galaxies. For the predictions of galaxy rotation curves and the galaxy cluster dynamics the best-fit values of $\alpha$ are determined by Eq. (\ref{alpha formula}).

The calculation of the matter power spectrum for $\rho_b >\rho_\phi$ will have unit oscillations. However, the finite size of galaxy survey samples and the associated window function used to produce presently available power spectra mask any such oscillations. Applying a window function to the MOG prediction for the matter power spectrum smooths out the power spectrum curve. The enhanced size of $G=G_N(1+\alpha)$ with $\alpha\sim 19$ predicts the right shape for the galaxy matter power spectrum curve, resulting in a fit to the data~\cite{MoffatToth3,MoffatToth4}. The GR prediction without dark matter and with $G=G_N$ cannot produce the correct magnitude or shape for the matter power spectrum. In Fig. 2, we show the predicted MOG matter power spectrum~\cite{MoffatToth3,MoffatToth4}.

In future galaxy surveys which utilize a large enough number of galaxies, with galaxies detected at sufficiently large redshift $z$, and with the use of a sufficiently narrow enough window function, it should be possible to detect any significant unit oscillations in the matter power spectrum, which can distinguish in the present universe between a dominant dark matter model and MOG without dark matter.

\
\begin{figure}
\centering \includegraphics[scale=0.75]{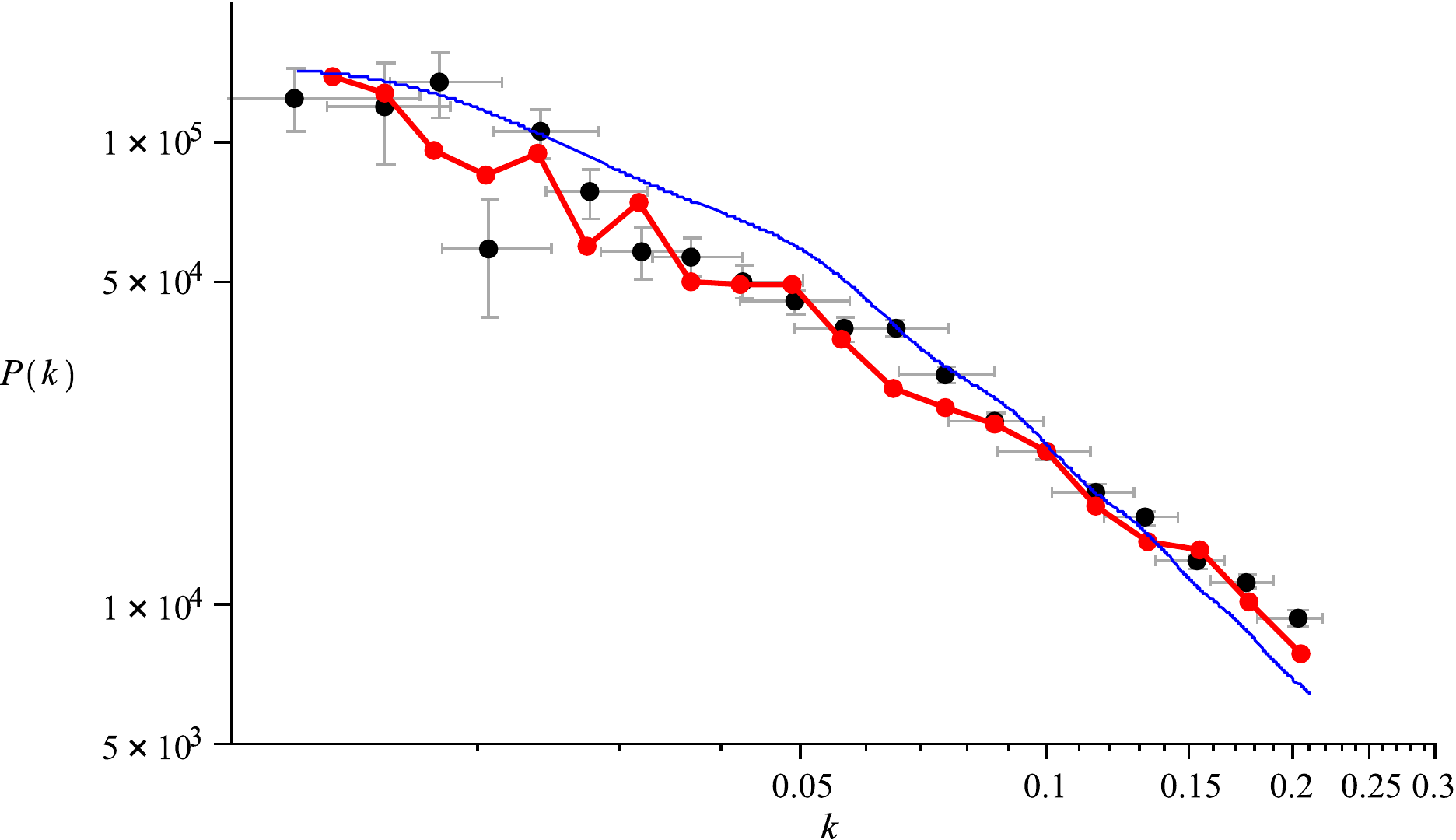}\\
\caption{\label{fig:matterpowerspectrum}After applying the appropriate window function, MOG (thick red line) shows agreement with the luminous red galaxy survey mass power spectrum data. The MOG fit to the data is comparable to the CDM prediction (thin blue line)~\cite{MoffatToth3,MoffatToth4}.}
\end{figure}

\section{Conclusions}

By assuming that the density of VGs associated with the massive and neutral vector field $\phi_\mu$ in MOG theory is the dominant density in the early universe with a VG mass, $m_\phi\sim 10^{-22}\,{\rm eV}$, the perturbations $\delta\rho_\phi$ satisfy the pressureless Jeans equation, allowing for an enhanced growth from the time of horizon entry and radiation-matter equality to produce large scale stellar and galaxy structure after the time of decoupling and recombination. The baryons obey the Jeans equation with baryon-photon pressure, so that the baryons oscillate between the time of horizon entry and decoupling, generating the baryon-pressure acoustical waves detected in the angular power spectrum in the CMB. The angular power spectrum calculation matches the $\Lambda$CDM model fit to the Planck collaboration 2018 data.

After the universe expands until the time of stellar and galaxy formation when $\rho_\phi < \rho_b$ and the mass of the VGs become ultra-relativistic with $m_\phi\sim 10^{-28}\,{\rm eV}$, the non-relativistic MOG Newtonian acceleration law, including the repulsive Yukawa interaction and the attractive gravitation due to the enhanced value of $G$, determines the rotation curves of galaxies and the dynamics of galaxy clusters.

The matter power spectrum in MOG can, with an appropriate window function, fit the galaxy matter power spectrum data. A critical test of MOG is whether significant baryon oscillations in the power spectrum begin to show as the number of observed luminous red galaxies increases and the size of the window function decreases. If the smooth ${\Lambda}CDM$ model fit to the matter power spectrum persists with a large enough increase in observed galaxies in galaxy redshift surveys, then this would rule out MOG in favor of the existence of dark matter halos in galaxies.

The ultra-light VG particles and their associated Proca field $\phi_\mu$ are an integral part of the MOG theory determined by the STVG action. The MOG theory can fit observational data from the solar system to galaxy and galaxy clusters, and the early universe cosmological data, and provides a gravitational description of the universe at both small and large distance scales.

\section{Appendix A}

\renewcommand{\theequation}{A-\arabic{equation}}
\newcommand{\alphaeqn}{\setcounter{saveeqn}{\value{equation}}%
\stepcounter{saveeqn}\setcounter{equation}{0}%
\renewcommand{\theequation}
 {\mbox{A-\arabic{saveeqn}\alph{equation}}}}%
\newcommand{\reseteqn}{\setcounter{equation}{\value{saveeqn}}
\renewcommand{\theequation}{A-\arabic{equation}}}
\setcounter{equation}{0}

The MOG action is given by
\begin{align}
S=S_G+S_\phi+S_S+S_M,
\end{align}
where $S_M$ is the matter action and
\begin{align}
S_G&=\frac{1}{16\pi}\int d^4x\sqrt{-g}\left[\frac{1}{G}(R+2\Lambda)\right],\\
\label{Baction}
S_\phi&=\int d^4x\sqrt{-g}\left[-\frac{1}{4}B^{\mu\nu}B_{\mu\nu}+\frac{1}{2}\mu^2\phi^\mu\phi_\mu\right],\\
S_S&=\int d^4x\sqrt{-g}\left[\frac{1}{G^3}\left(\frac{1}{2}g^{\mu\nu}\partial_\mu G\partial_\nu G-V_G\right)+\frac{1}{\mu^2G}\left(\frac{1}{2}g^{\mu\nu}\partial_\mu\mu\partial_\nu\mu-V_\mu\right)\right],
\end{align}
where
$B_{\mu\nu}=\partial_\mu\phi_\nu-\partial_\nu\phi_\mu$ and $V_G$ and $V_\mu$ are potentials.  Note that we choose units such that
$c=1$, and use the metric signature $[+,-,-,-]$. The Ricci tensor is
\begin{equation}
R_{\mu\nu}=\partial_\lambda\Gamma^\lambda_{\mu\nu}-\partial_\nu\Gamma^\lambda_{\mu\lambda}
+\Gamma^\lambda_{\mu\nu}\Gamma^\sigma_{\lambda\sigma}-\Gamma^\sigma_{\mu\lambda}\Gamma^\lambda_{\nu\sigma}.
\end{equation}

The matter stress-energy tensor is obtained by varying the
matter action $S_M$ with respect to the metric:
\begin{align}
T^{\mu\nu}_M=-2(-g)^{-1/2}\delta S_M/\delta g_{\mu\nu}\,.
\end{align}
Varying $S_\phi$ and $S_S$ with respect to the metric yields
\be
T^{\mu\nu}_\phi=-2(-g)^{-1/2}[\delta S_\phi/\delta g_{\mu\nu}],
\ee
\be
T^{\mu\nu}_S=-2(-g)^{-1/2}[\delta S_S/\delta g_{\mu\nu}].
\ee
Combining these gives the total stress-energy tensor
\be
T^{\mu\nu}=T^{\mu\nu}_{\rm M}+T^{\mu\nu}_\phi+T^{\mu\nu}_S.
\ee
We have
\be
T_{\phi\mu\nu}=-\frac{1}{4\pi}\bigg[B^\alpha_\mu B_{\nu\alpha}-g_{\mu\nu}\bigg(B^{\rho\alpha}B_{\rho\alpha}+\frac{1}{2}\mu^2\phi^\mu\phi_\mu\bigg)\bigg].
\ee

The MOG field equations are given by
\begin{align}
G_{\mu\nu}-\Lambda g_{\mu\nu}+Q_{\mu\nu}=8\pi GT_{\mu\nu}\,,
\label{eq:MOGE}
\end{align}
\begin{equation}
\label{Bequation1}
\frac{1}{\sqrt{-g}}\partial_\mu\biggl(\sqrt{-g}B^{\mu\nu}\biggr)+\mu^2\phi^\nu=J^\nu,
\end{equation}
\be
\label{Bequation2}
\partial_\sigma B_{\mu\nu}+\partial_\mu B_{\nu\sigma}+\partial_\nu B_{\sigma\mu}=0.
\ee
Here, $G_{\mu\nu}=R_{\mu\nu}-\frac{1}{2} g_{\mu\nu} R$
is the Einstein tensor and
\begin{align}
  Q_{\mu\nu}=\frac{2}{G^2}(\partial^\alpha G \partial_\alpha G\,g_{\mu\nu}
  - \partial_\mu G\partial_\nu G) - \frac{1}{G}(\Box G\,g_{\mu\nu}
  - \nabla_\mu\partial_\nu G)
  \label{eq:Q}
\end{align}
is a term resulting from the the presence of second derivatives
of $g_{\mu\nu}$ in $R$ in $S_G$. We also have the field equations:
\begin{equation}
\Box G=K,\quad\Box\mu=L,
\end{equation}
where $\nabla_\mu$ is the covariant derivative with respect to the metric $g_{\mu\nu}$, $\Box=\nabla^\mu\nabla_\mu$, $K=K(G,\mu,\phi_\mu)$ and $L=L(G,\mu,\phi_\mu)$.

Combining the Bianchi identities, $\nabla_\nu G^{\mu\nu}=0$, with the
field equations (\ref{eq:MOGE}) yields the conservation law
\begin{align}
  \nabla_\nu T^{\mu\nu}+\frac{1}{G}\partial_\nu G\,T^{\mu\nu} -
  \frac{1}{8\pi G}\nabla_\nu Q^{\mu\nu}=0 \,.
  \label{eq:Conservation}
\end{align}

It is a key premise of MOG that all baryonic matter possesses, in
proportion to its mass $M$, positive gravitational charge:
$Q_g=\kappa\,M$.  This charge serves as the source of the vector field
$\phi^\mu$.  Moreover, $\kappa=\sqrt{8\pi(G-G_N)}=\sqrt{8\pi\alpha G_N}$, where
$G_N$ is Newton's gravitational constant and
$\alpha=(G-G_N)/G_N\ge 0$.  Variation of $S_M$ with respect to the
vector field $\phi^\mu$, then yields the MOG 4-current
$J_\mu=-(-g)^{-1/2}\delta S_M/\delta \phi^\mu$.

For the case of a perfect fluid:
\begin{equation}
\label{energymom}
T^{\mu\nu}=(\rho+p)u^\mu u^\nu-pg^{\mu\nu},
\end{equation}
where $\rho$ and $p$ are the matter density and pressure, respectively, and $u^\mu$ is the 4-velocity of an element of the fluid.
We obtain from (\ref{energymom}) and $u^\mu u_\mu=1$ the 4-current:
\begin{equation}
\label{current}
J_\mu=\kappa T_{M\mu\nu}u^\nu=\kappa\rho u_\mu.
\end{equation}
It is shown in~\cite{Roshan1} that, with the assumption $\nabla_\mu J^\mu=0$,
(\ref{eq:Conservation}) reduces to
\begin{align}
\nabla_\nu T_M^{\mu\nu}=B_\nu^{~\mu} J^\nu\,.\label{eq:Mcons}
\end{align}

The field equations (\ref{Bequation1}) and (\ref{Bequation2}) are to be considered as effective field equations for the $B_{\mu\nu}$ field. The scalar field $G$ is in effect a running coupling constant dependent on the binding energy or mass of a body. This also applies to the scalar field $\alpha$. The field equations for $B_{\mu\nu}$ do not solve to give unique universal values of $\alpha$ or $\mu$. The solutions for $G$ or $\alpha$ depend on the renormalization group value of $G$ analogous to the behavior of coupling constants in the standard model of particle physics.

The equation of motion of a particle are given by
\be
\frac{d^2x^\mu}{ds^2}+{\Gamma_{\alpha\beta}}^\mu u^\alpha u^\beta=\frac{q_g}{m}{B^\mu}_\alpha u^\alpha,
\ee
where $q_g=\kappa m$ and $m$ is the particle mass. We have $q_g/m=\kappa$ and the equation of motion satisfies the equivalence principle. Due to the Lorentz-type gravitational force, massive particles do not free fall on geodesics. Photons have $m_\gamma=0$ and satisfy the null geodesic equation:
\be
k^\mu\nabla_\mu k^\nu = 0,
\ee
where $k^\mu$ is the 4-momentum vector of photons and $k^2=g^{\mu\nu}k_\mu k_\nu=0$. Gravitational radiation follows the same null geodesics as photons~\cite{GreenMoffatToth,Liu}, and gravitational waves travel with the speed of light.

For the low energy behavior of $G=G_N(1+\alpha)$ and $\mu$, we can adopt approximately constant values for $G$ and $\mu$: $\partial_\mu G\approx 0$ and $\partial_\mu\mu\approx 0$, where $\alpha$ depends on the density and mass of a body. A phenomenological formula for $\alpha$ is:
\be
\label{alpha formula}
\alpha=\alpha_{\inf}\frac{M}{(\sqrt{M}+E)^2},
\ee
where $\alpha={\cal O}(10)$ and $E=2.5\times 10^4\,M_{\odot}$. For $r\ll \mu^{-1}\sim 25 -100$ kpc the MOG acceleration (\ref{MOG acceleration}) reduces to the Newtonian acceleration ${\vec a}_{\rm Newt}$. This is consistent with $\alpha = 10^{-9}$ obtained from (\ref{alpha formula}) for $M=1\, M_{\odot}$, guaranteeing that the solar system observational data is satisfied.

\section*{Acknowledgments}

I thank Martin Green and Viktor Toth for helpful discussions. Research at the Perimeter Institute for Theoretical Physics is supported by the Government of Canada through industry Canada and by the Province of Ontario through the Ministry of Research and Innovation (MRI).

\end{document}